\documentclass[prb,twocolumn,groupedaddress]{revtex4}

\usepackage{graphicx}
\usepackage{dcolumn}
\usepackage{bm}

\begin{document}

\preprint{ATB-1}

\title{Stripe Order and Magnetic Transitions in La$_{2-x}$Sr$_{x}$NiO$_4$}

\author{P. G. Freeman}
\homepage{http://xray.physics.ox.ac.uk/Boothroyd}
\author{A. T. Boothroyd}
\author{D. Prabhakaran}
\affiliation{Department of Physics, Oxford University, Oxford, OX1 3PU, United
Kingdom }

\author{M. Enderle}\affiliation{
Institut Laue-Langevin, BP 156, 38042 Grenoble Cedex 9, France }

\author{C. Niedermayer}\affiliation{
Laboratory for Neutron Scattering, ETHZ and PSI, CH-5232 Villigen
PSI, Switzerland }

\date{\today}

\begin{abstract}

Magnetic order has been investigated in stripe-ordered
La$_{2-x}$Sr$_{x}$NiO$_4$ ($x = 0.275, 0.37, 0.4$) by d.c.
magnetization and by polarized- and unpolarized-neutron
diffraction. In the magnetically ordered phase, all three
compositions exhibit a magnetic transition consistent with a spin
reorientation in the $ab$ plane. For $x = 0.37$, the spin axes
rotate from an angle of 37.7 $\pm$ 0.3$^{\circ}$ to the stripe
direction at $71\ $K, to 52.3 $\pm$ 0.2$^{\circ}$ at $2\ $K. The
$x = 0.275$ and $0.4$ compounds were found to undergo a similar
spin reorientation. A spin reorientation has now been observed to
occur for five different doping levels in the range $0.275\leq\ x
\leq\ 0.5$, suggesting that this spin transition is an intrinsic
property of the stripe phase.

\end{abstract}

\maketitle

\section{\label{sec:intro}Introduction}

La$_{2-x}$Sr$_x$CuO$_4$(LSCO) and La$_{2-x}$Sr$_x$NiO$_4$(LSNO)
are isostructural, but it is well known that LSCO
superconducts\cite{Bednorz} when sufficiently doped whereas LSNO
does not. In LSNO the doped charges are known to localize in the
form of charge stripes, i.e. periodically spaced lines of charges
at $45^{\circ}$ to the Ni-O bonds. Antiferromagnetic ordering of
the Ni spins between the charge stripes sets in at lower
temperatures.\cite{yoshizawa-PRB-2000} The charge-stripes act as
antiphase domain walls to the antiferromagnetic background. The
pattern of incommensurate magnetic fluctuations in
LSCO\cite{super} resembles the magnetic order seen in LSNO, and
this has been attributed to the existence of dynamic stripes in
LSCO. The fluctuations in superconducting LSCO are centred on
wavevectors parallel to the Cu-O bonds,\cite{yamada-PRB-2002}  but
the fluctuations in the non-superconducting state, $0.024 \leq x
\leq  0.053$, are centred on wavevectors at $45^{\circ}$ to the
Cu-O bonds.\cite{superstripe} These parallels suggest that charge
stripe correlations may play an important role in superconducting
LSCO. \cite{tranquada-Nature-1995}

Charge-stripe order has been studied in LSNO by
neutron\cite{neutron,yoshizawa-PRB-2000,lee-PRB-2001,kajimoto-PRB-2001,kajimoto,
me} and x-ray\cite{x-ray,pash-PRL-2000,HATTON} diffraction for
doping levels in the range 0.135 $\leq x \leq$ 0.5. As well as
being static on the time scale probed by neutrons and x-rays, the
charge stripes are found to be well correlated with correlation
lengths in excess of 100\,\AA\ for certain levels of
doping.\cite{yoshizawa-PRB-2000,pash-PRL-2000,HATTON} These two
properties make LSNO a good system for studying the basic
properties of spin-charge stripes.

\begin{figure}[!ht]
\begin{center}
\includegraphics{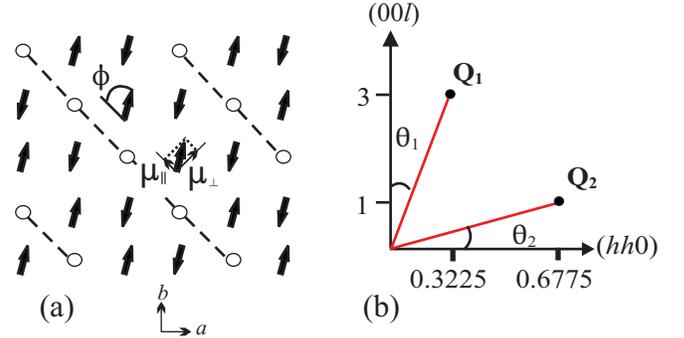}
\caption[Stripe-order and peak positions]{(a) Model for the stripe
pattern ordering in the $ab$ plane of La$_{5/3}$Sr$_{1/3}$NiO$_4$.
Circles represent holes residing on Ni$^{3+}$ sites, while the
solid arrows represent the Ni$^{2+}$ spins. Dashed lines indicate
the charge domain walls of the stripe-order. The occurence of
purely Ni centred stripes may not be realized in practice,
especially for $x \neq 1/3$. The observed spin order in
La$_{1.63}$Sr$_{0.37}$NiO$_{4+\delta}$ is similar to that shown,
but is actually incommensurate in the direction perpendicular to
the stripes. The components of the ordered moment parallel
($\mu$$_\|$) and perpendicular ($\mu$$_\bot$) to the stripe
direction are shown, and $\phi$ denotes the angle between the spin
axis and the stripe direction. (b) Diagram of the ($h,\ h,\ l$)
plane in reciprocal space. {\bf Q$_1$} and {\bf Q$_2$} are the
scattering vectors of two magnetic Bragg peaks of the stripe
order, chosen to be approximately parallel and perpendicular to
the c axis. The particular positions shown in the diagram are
those investigated for the case $x = 0.37$.}
\label{fig:Figure1Sr=0_37mk2}
\end{center}
\end{figure}

These studies have revealed many key facts about charge-stripe
ordering including the variation of the stability of
charge-ordering with doping. LSNO with $x$ = 1/3 has stripe order
that is particularly stable owing to a commensurability effect
that pins the charges to the
lattice.\cite{ramirez-PRL-1996,yoshizawa-PRB-2000,kajimoto-PRB-2001}
Figure \ref{fig:Figure1Sr=0_37mk2}(a) shows the commensurate
charge-ordering that occurs for $x$ = 1/3. Although this figure
shows the charge stripes only residing on the Ni sites, recent
tunnelling electron microscopy work has shown the charge stripes
can also reside on the O sites.\cite{Li-PRB-2003} The charge order
is even more stable for $x = 1/2$, forming a `checkerboard'
pattern in the Ni-O$_2$ layers at temeratures below $\sim450\ K$,
which becomes slightly incommensurate below $\sim180\
K$.\cite{chen-PRL-1993,kajimoto}

Lee {\it et al.}\cite{lee-PRB-2001} studied the magnetic order in
LSNO crystals with $x = 0.275$ and $1/3$ with polarized neutrons
in order to determine the direction of the ordered moment. They
concluded that at $T = 11\ $K the spins in the Ni-O$_2$ layers are
aligned at an angle $\pm\  \phi$ to the stripe direction, where
$\phi = 27^{\circ}$ for $x = 0.275$ and $52.5^{\circ}$ for $x =
1/3$. However, for the $x = 1/3$ they found a reorientation
transition at $T_{\rm SR} \simeq 50\ $K such that on warming the
spins rotate by an angle of $\Delta\phi = 12.5^{\circ}$ towards
the stripe direction. Freeman {\it et al.}\cite{me} studied a
sample with $x=1/2$ and observed a similar spin reorientation
transition: in this case at $T = 2\, K$ $\phi = 78^{\circ}$,
$T_{\rm SR} \simeq 57\ $K, and $\Delta\phi = 27^{\circ}$.

Although this much is known about the spin orientation in LSNO the
trends have not yet been fully established, and the mechanism
driving the spin reorientation is not understood. Our new study
was undertaken to try to address these questions by studying
doping levels other than $1/3$ and $1/2$.

We studied single crystals of La$_{2-x}$Sr$_{x}$NiO$_{4}$ grown by
the floating-zone method,\cite{Prab} using the techniques of
magnetometry, ($x = 0.275, 0.37, 0.4$), polarized-neutron
diffraction ($x = 0.37$) and unpolarized-neutron diffraction ($x =
0.275, 0.4$). The charge and magnetic ordering temperatures were
found to be in good agreement with previous work on samples of
similar doping.\cite{yoshizawa-PRB-2000,kajimoto} The data reveal
a spin reorientation similar in size and orientation to that in
La$_{5/3}$Sr$_{1/3}$NiO$_{4}$, but which is slower and occurs at
lower temperatures, $\sim 10\, K$, for all three doping levels
studied. These spin reorientations, unlike those in the $x = 1/2$
or $1/3$ doped materials, all occur for incommensurate doping
levels.

\section{\label{sec:exper}Magnetization Measurements}

Magnetization data were collected with a superconducting quantum
interference device (SQUID) magnetometer (Quantum Design), with
the field applied parallel to the $ab$ plane of the crystal. The
crystals used for the magnetization measurements had typical
dimensions $\sim$$5\times5\times2$\,mm$^{3}$. We carried out dc
measurements either by cooling the sample in an applied field of
500 Oe parallel to the $ab$ plane (FC), or by cooling in zero
field then measuring while warming in a field of 500 Oe (ZFC).

\begin{figure}[!ht]
\begin{center}
\includegraphics[width=8cm,clip=]{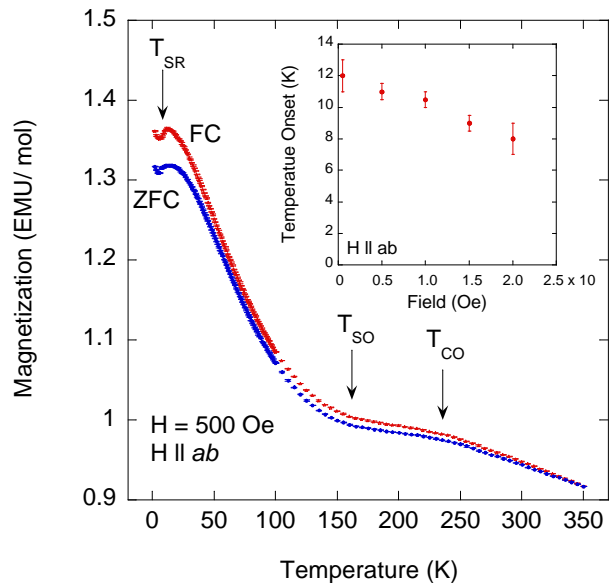}
\caption[SQUID]{FC and ZFC magnetization data for
La$_{1.63}$Sr$_{0.37}$NiO$_{4+\delta}$. Arrows indicate the
charge-ordering temperature, $T_{\rm CO}$, spin-ordering
temperature, $T_{\rm SO}$, and the spin-reorientation temperature,
$T_{\rm SR}$, determined separately by neutron diffraction. The
inset shows the field dependence of the onset temperature of the
spin reorientation.} \label{fig:SQUID}
\end{center}
\end{figure}

In figure \ref{fig:SQUID} we show the variation of the FC and ZFC
magnetizations for $x = 0.37$. A subtle change of slope at $240\pm
10$\,K, marks the charge-ordering temperature, with a more
pronounced gradient change at $160\pm 10$\,K marking the
spin-ordering temperature. We observe from these results that this
material exhibits irreversible magnetic behaviour, with a large
FC--ZFC difference below $\sim 50$\,K and a much smaller
difference that persists to the charge-ordering temperature
$T_{\rm CO} \sim 240$\,K with a slight widening around the
magnetic ordering temperature, $T_{\rm SO} \sim 180$\,K. This is a
typical feature for magnetization results on LSNO compounds that
will be reported in detail elsewhere.\cite{unpublished} Of most
concern to the present work is the small but sharp drop in
magnetization observed at $T_{\rm SR}$ $\sim 10$\,K. The inset of
figure \ref{fig:SQUID} shows the field dependence of $T_{\rm SR}$,
which can be seen to decrease when increasing the applied field.

\begin{figure}[!ht]
\begin{center}
\includegraphics[width=8cm,clip=]{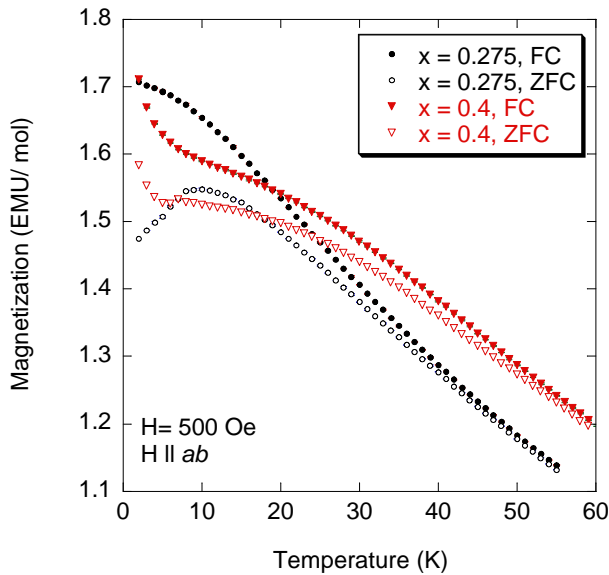}
\caption[Squid]{FC and ZFC magnetization data for
La$_{2-x}$Sr$_{x}$NiO$_{4+\delta}$ $ x = 0.275$ and $0.4$.}
\label{fig:SQUID2}
\end{center}
\end{figure}

Figure \ref{fig:SQUID2} shows the variation of FC and ZFC
magnetizations for the $x = 0.275$ and $x = 0.4$. Like $x = 0.37$,
both these materials are observed to have irreversible magnetic
behaviour. The ZFC magnetization of the $x = 0.275$ crystal has a
rounded maximum at $\sim 10$\,K. For $x = 0.4$ there is no
maximum, but the increase in magnetization with decreasing
temperature first slows down and then begins to rise sharply below
5\,K. The increase below 5\,K could be due to a small amount of
paramagnetic impurity in the crystal.

\section{\label{sec:exper}Neutron Diffraction Measurements}

The polarized neutron experiments were performed on the
triple-axis spectrometer IN20 at the Institut Laue-Langevin. The
energies of the incident and elastically scattered neutrons were
selected by Bragg reflection from an array of Heusler alloy
crystals. The data were obtained with initial and final neutron
wavevectors of 2.66\,\AA$^{-1}$. A PG filter was present between
the sample and the analyzer to suppress scattering of higher-order
harmonics.  The unpolarized neutron experiments were performed on
the triple-axis spectrometer RITA-II at SINQ at the Paul Scherrer
Institut. The energies of the incident and elastically scattered
neutrons were selected by Bragg reflection from a PG crystal. The
data were obtained with initial and final neutron wavevectors of
1.55\,\AA$^{-1}$, and a Be filter operating at 77\,K was present
between the sample and the analyzer to suppress scattering of
higher-order harmonics.

For $x = 0.275, 0.37$, single crystal rods of 7--8\,mm diameter
and $\sim$45\,mm in length were used, and for $x = 0.4$ the
crystal was a slab of dimensions
$\sim$$15\times10\times4$\,mm$^{3}$. In this work we  describe the
structural properties of LSNO with reference to a tetragonal unit
cell, with unit cell parameters $a \approx 3.8\AA$, $c \approx
12.7\AA$.  The samples were mounted with the [001] and [110]
crystal directions in the horizontal scattering plane. Scans were
performed in reciprocal space either parallel to the ($h$,\ $h$,\
0) direction at constant $l$, or parallel to the (0,\ 0,\ $l$)
direction at constant $h$.

We begin by discussing the polarized-neutron diffraction results
for $x = 0.37$. Initially, the neutron polarization ${\bf P}$ was
arranged to be parallel to the scattering vector ${\bf Q}$, by an
adjustable guide field of a few mT at the sample position. In this
configuration a neutron's spin is flipped during an interaction
with electronic magnetic moments, but remains unchanged when
scattered by a non-magnetic process, e.g. a lattice distortion.
Thus by measuring the spin-flip (SF) and non-spin-flip (NSF)
channels one can identify whether observed scattering is magnetic
or not in origin.

\begin{figure}[!ht]
\begin{center}
\includegraphics{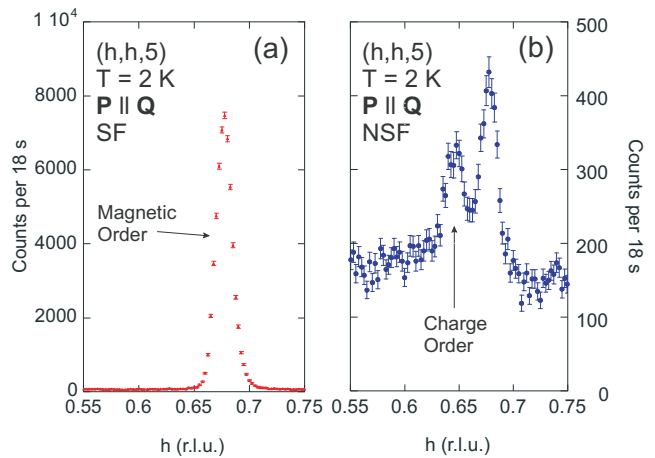}
\caption[hhscan]{(a) The spin flip (SF) diffraction channel for a
scan parallel to $(h,\ h,\ 0)$ for $l$ = 5 for
La$_{1.63}$Sr$_{0.37}$NiO$_4$ at $T = 2$\, K. This peak
corresponds to the magnetic order peak, and is centred at $h =
0.678$. (b) The non-spin flip diffraction for the same scan. The
arrow indicates the charge-order Bragg peak. The second peak is
diffraction from the magnetic order peak observed in the NSF
channel due to the imperfect spin polarization of the neutron
beam.} \label{fig:hh5}
\end{center}
\end{figure}

Magnetic order Bragg peaks were observed at
($h+1/2\pm\varepsilon/2$, $h+1/2\pm\varepsilon/2$,\ $l$) positions
for all integer $l$. This can be seen in figure \ref{fig:hh5}(a),
which shows the SF scattering for a scan parallel to ($h,\ h,$\ 0)
for $l$ = 5 at $T = 2$\,K.  The peak positions corresponds to
$\varepsilon = 0.3554 \pm 0.0002$, consistent with previous
measurements. \cite{yoshizawa-PRB-2000}

Figure\ \ref{fig:hh5}(b) shows the NSF scattering for the same
scan as Fig.\ \ref{fig:hh5}(a). The scan contains 2 weak peaks,
one at $h =$ 0.646 $\pm$ 0.001 corresponding to charge ordering
with an incommensurability of $\varepsilon$ = 0.354 $\pm$ 0.001,
and the other at $h =$ 0.678 corresponding to magnetic ordering.
The latter appears in the NSF channel due to imperfect
polarization of the neutron beam. We searched for the charge order
peak at other equivalent $(0.646,\ 0.646,\ l)$ positions, but only
at $l =$ 3 and 5 was there a measurable peak. From the temperature
dependence of the charge peak in Fig.\ \ref{fig:hh5}(b) we found
$T_{\rm CO} \approx 230 $K. By performing scans parallel to $(h,\
h,\ 0)$ we were able to obtain the in-plane charge-order
correlation length perpendicular to the stripe direction of 70
$\pm$ 6\,\AA. This compares with a correlation length of 49 $\pm$
5\,\AA\ along the $c$ axis. These results show that the charge
order is relatively three-dimensional.

\begin{figure}[!ht]
\begin{center}
\includegraphics[width=8cm,clip=]{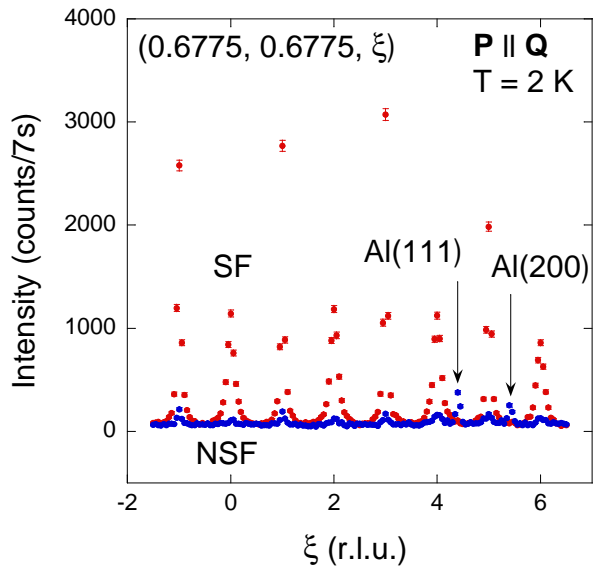}
\caption[lscan]{Spin-flip (SF) and non-spin-flip (NSF) diffraction
from La$_{1.63}$Sr$_{0.37}$NiO$_4$ at $T$ = 2\,K. The scan is
parallel to (0,\ 0,\ $l$) and passes through the magnetic-order
peak (0.6775,\ 0.6775,\ 1). No correction has been made for the
imperfect polarization of the neutron beam. The additional peaks
in the NSF channel at $l \approx$ 4.4 and 5.4 are due to
diffraction from the Al sample mount.} \label{lscan}
\end{center}
\end{figure}

Figure\ \ref{lscan} shows a scan parallel to $(0,\ 0,\ l)$ through
the magnetic order peaks. The widths of the peaks in this scan
relate to the correlation length along the c-axis, however we
observed that the correlation lengths for even and odd $l$ differ
by a factor $\approx$ 2. That is, for even $l$ we obtain a
correlation length of 53.2 $\pm$ 1.4\,\AA\ and for odd $l$ we
obtain a correlation length 108 $\pm$ 2\,\AA\ . We performed scans
parallel to $(h,\ h,\ 0)$ on odd $l$ peaks, for which we obtained
an in-plane correlation length perpendicular to charge stripe
direction of 112.6 $\pm$ 1.1\,\AA\ .

The intensities of the even and odd $l$ magnetic peaks were
discussed in work on La$_{2}$NiO$_{4+\delta}$, by P. Wochner {\it
et al.}.\cite{wochner-PRB-1998} For a commensurate stripe
spin-ordering, such as $\varepsilon = 1/3$, the stripes stack in a
body centred arrangement and only the odd $l$ magnetic peaks are
observed, with the systematic absence of the $l =$ even peaks.
However, for incommensurate spin-ordering with the stripes either
pinned to the Ni or O sites,\cite{Li-PRB-2003} perfect
body-centred stacking cannot be achieved. The disorder thus
created, along with the additional disorder introduced due to
differing Coulomb interactions between the $ab$ layers, result in
the presence of $l =$ even peaks. Hence, $l =$ odd peaks have a
long correlation length as they come from the ideal long range
body-centred stacking, whereas $l =$ even peaks are a result of
the disorder created on the smaller length scale of the non-ideal
stacking.

\begin{figure}[!ht]
\begin{center}
\includegraphics{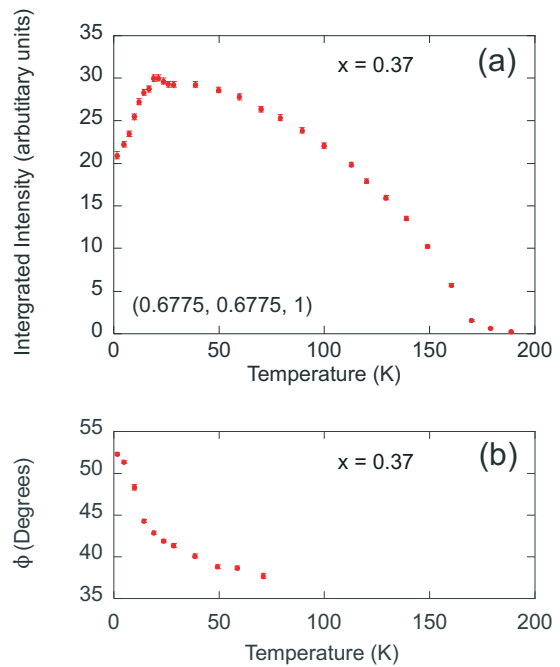}
\caption[peak intensity and spin direction]{(a) The temperature
dependence of the intensity of the magnetic Bragg peak at ${\bf
Q}_2=(0.6775,0.6775,1)$
--- see figure 1(b). (b) The temperature
dependence of the angle $\phi$ between the spin axis and the
stripe direction obtained from polarized-neutron analysis. }
\label{fig:polar}
\end{center}
\end{figure}

In figure\ \ref{fig:Figure1Sr=0_37mk2}(b) are shown the positions
of two magnetic reflections {\bf Q$_1$} = (0.3225,0.3225,3) and
{\bf Q$_2$} = (0.6775,0.6775,1). The vectors {\bf Q$_1$} and {\bf
Q$_2$} are directed close to $(0,\ 0,\ l)$ and $(h,\ h,\ 0)$,
respectively. Since magnetic neutron diffraction is sensitive to
spin components perpendicular to {\bf Q}, the scattering at {\bf
Q$_1$} arises mainly from the total in-plane spin moment, while
that at {\bf Q$_2$} comes mainly from the spin components parallel
to the stripe direction and along the c axis. Hence, we performed
scans parallel to $(h,\ h,\ 0)$ through {\bf Q$_2$} at different
temperatures to give a first indication as to whether there exists
an in-plane spin reorientation like those observed for $x = 1/3$
and 1/2. Figure\ \ref{fig:polar}(a) shows the temperature
dependence of the magnetic reflection {\bf Q$_2$}. The magnetic
ordering transition can be seen to occur at $T_{\rm SO}$ $\simeq$
170\,K. On cooling below $T_{\rm SO}$ the intensity of {\bf Q$_2$}
can be seen to increase monotonically until it reaches a maximum
at $\simeq$ 20\,K, then it is seen to decrease in intensity
continuously to our base temperature. This anomalous behaviour
correlates well with the transition observed in the magnetization,
Fig. \ref{fig:SQUID}, and indicates a spin reorientation below
$\simeq$ 20\,K.

To fully analyze the direction of the spins over this temperature
range we varied the direction of the neutron polarization {\bf P}
relative to the scattering vector {\bf Q}, measuring at the ${\bf
Q}_1$ and ${\bf Q}_2$ positions. The method is described in Ref
\onlinecite{me}, where the expressions used to obtain the spin
direction are given. A correction for the slightly non-ideal
performance of the polarization elements of the instrument was
calculated from the flipping ratio of 18 $\pm$ 1 measured on the
magnetic Bragg peaks.

Polarization analysis of both the {\bf Q$_1$} and the {\bf Q$_2$}
Bragg peaks revealed that the moment lies in the $ab$ plane to
within 1$^{\circ}$ in the temperature range $2-71$\,K. Having
established this, we subsequently assumed the $c$-axis component
to be zero and analyzed the polarization at {\bf Q$_1$} to
determine the in-plane moment. From this analysis we determined
that the spins rotated from an angle of 37.7 $\pm$ 0.3$^{\circ}$
to the stripe direction at $T$ = 71\,K to 52.3 $\pm$ 0.2$^{\circ}$
at $T$ = 2\,K, as shown in Fig. \ref{fig:polar}(b). The transition
occurs mainly between 10 and 20\,K, but slowly develops from below
$\sim 50$\,K.

\begin{figure}[!ht]
\begin{center}
\includegraphics[width=8cm,clip=]{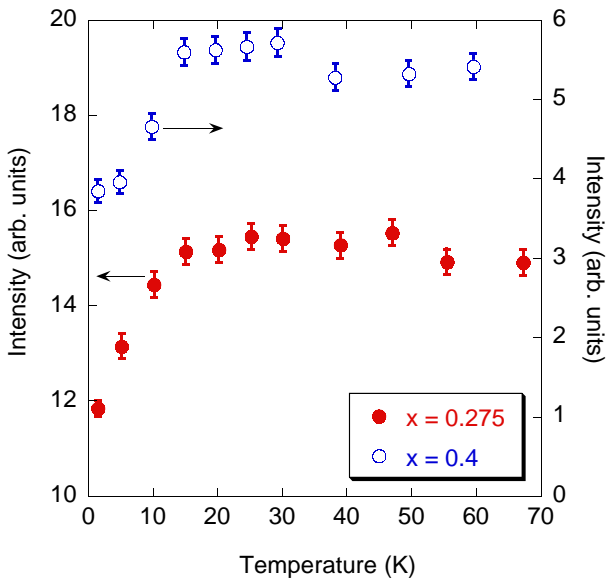}
\caption[peak intensity and spin direction]{The temperature
dependence of the intensity of the magnetic Bragg peak at ${\bf
Q}_2=(0.645,0.645,0)$ for $x$ = 0.275, and at ${\bf
Q}_2=(0.685,0.685,1)$ for $x$ = 0.4.} \label{fig:RITAII}
\end{center}
\end{figure}

We performed unpolarized neutron diffraction at equivalent {\bf
Q$_1$} and {\bf Q$_2$} positions on the single crystals with $x =
0.275$ and $x = 0.4$, using the instrument RITA-II at SINQ. We
found a similar behaviour to the $x = 0.37$ results just
described. In particular, the temperature dependence of the {\bf
Q$_2$} = (0.645,0.645,0) magnetic Bragg peak for $x = 0.275$, and
of the {\bf Q$_2$} = (0.685,0.685,1) magnetic Bragg peak for $x =
0.4$, both have a maximum similar to that shown in Fig.\
\ref{fig:polar}(a) for $x = 0.37$. By contrast, the intensities of
the {\bf Q$_1$} Bragg peaks are almost constant below 20\, K. The
temperature dependence of the {\bf Q$_2$} intensity is shown on
Fig.\ \ref{fig:RITAII} for both $x = 0.275$ and $x = 0.4$. The
drop in {\bf Q$_2$} intensity at low temperature implies a spin
reorientation in $x = 0.4$ and $x = 0.275$ similar in nature to
that in $x = 0.37$.

Unpolarized neutron diffraction cannot accurately determine $\phi$
without a detailed analysis of the intensities of many diffraction
peaks, but we can estimate $\Delta\phi$ from the drop in intensity
of the {\bf Q$_2$} peak below $T_{\rm SR}$ $\sim\ 10$\,K and the
value of $\phi$ for $T >T_{\rm SR}$, assuming the ordered moment
remains in the $ab$ plane and fixed in magnitude in this
temperature range. Taking $\phi = 27^{\circ}$ above $T_{\rm SR}$
for $x = 0.275$,\cite{lee-PRB-2001} and using $\phi= 38^{\circ}$
for $x = 0.4$ (based on the observations for $x = 0.37$ at $71\,
K$). We find $\Delta\phi$ $\approx 10-15^{\circ}$ for both $x =
0.275$ and $x= 0.4$, similar to $x = 0.37$.

\section{\label{sec:conc}Discussion and Conclusions}

\begin{figure}[!ht]
\begin{center}
\includegraphics{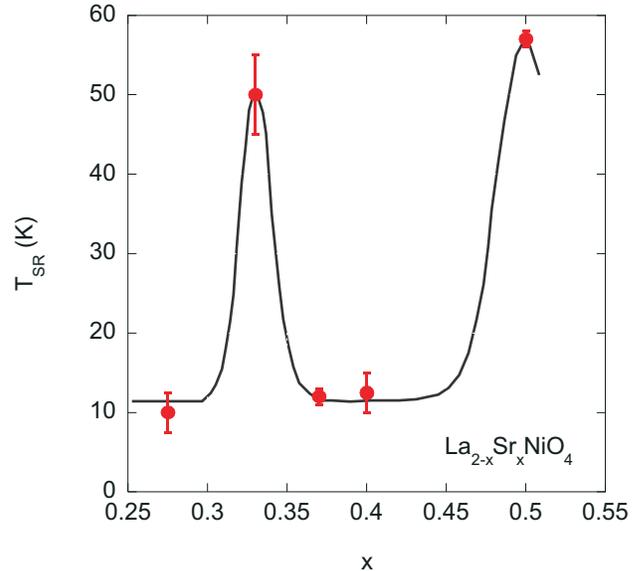}
\caption[order temperature]{Variation of the spin reorientation
temperature, $T_{\rm SR}$, with doping level, $x$. The line is a
guide to the eye indicating the possible trend.} \label{fig:SR}
\end{center}
\end{figure}

There are differences and similarities between the spin
reorientations reported here for $x$ = 0.275, 0.37 and $ 0.4$ and
those that occur for $x$ = 1/3,\cite{lee-PRB-2001} and $x =
1/2$.\cite{me} In each case the spins rotate in the same sense,
away from the stripe direction on cooling. The size of the
reorientation in $x$ = 0.37 ($\Delta\phi$ $\simeq 14.5^{\circ}$)
(and more approximately in $x = 0.275$ and $0.4$) is similar to
that in $x$ = 1/3 ($\Delta\phi$ $\simeq 13^{\circ}$), but smaller
than in $x$= 1/2 ($\Delta\phi$ $\simeq 26^{\circ}$). However, in
$x = 0.37, 0.275$ and $0.4$ the spin reorientation occurs at a
much lower temperature, $T_{\rm SR}$$\simeq 10$\,K, compared with
$T_{\rm SR}$ $\simeq 50$\,K for $x$ = 1/3 and $T_{\rm SR}$ $\simeq
57$\,K for $x$ = 1/2. Figure \ref{fig:SR} summarises the variation
of $T_{\rm SR}$ with $x$ that has so far been established for the
doping range  $0.275\leq\ x \leq\ 0.5$. The results indicate that
if there is no commensurate charge ordering the spin reorientation
occurs $\sim 10$\,K, but if there is a commensurate charge
ordering the spin reorientation occurs at $\sim 50$\,K.

There is also evidence of a trend in the direction of the ordered
moment.  The base temperature spin orientations are  $53^{\circ}$
for $x$= 1/3, $52^{\circ}$ for $x = 0.37$, and $78^{\circ}$ for
$x$= 1/2, with charge ordering temperatures of  $\sim 240$\,K,
$\sim 230$\,K, $\sim 480$\,K respectively. We can add to this list
an estimate of $38-43^{\circ}$ for $x = 0.275$, (charge ordering
temperature $\sim 200\, K$) based on the angle of $27^{\circ}$ at
11\, K found by Lee {\it et al.} \cite{lee-PRB-2001}, with an
additional $10-15^{\circ}$ due to the spin reorientation on
cooling to 2\,K. Hence, there seems to be a correlation between
the spin orientation angle $\phi$ and the charge-ordering
temperature. Further measurements on samples with doping levels
between $x = 0.4$ and 0.5 would be useful to confirm this trend.

Up to now, spin reorientations in La$_{2-x}$Sr$_{x}$NiO$_{4}$ had
only been observed in materials with especially stable charge
order ($x = 1/3$ and 1/2). The existence of a spin reorientation
in $x$ = 1/2 has shown that a commensurate spin-stripe order is
not required for a spin reorientation to occur.\cite{me} The new
results on $x$ = 0.275, 0.37 and 0.4 presented here further show
that not even commensurate doping is required. It is likely that
LSNO at all doping levels in the range $0.275\leq\ x \leq\ 0.5$
undergo a spin reorientation, but that $T_{\rm SR}$ is larger at
the commensurate doping compositions.

This work was supported in part by the Engineering and Physical
Sciences Research Council of Great Britain. Some of this work was
performed at the Swiss Spallation Neutron Source SINQ, Paul
Scherrer Institute (PSI), Villigen, Switzerland.

\end{document}